\documentclass[aps,showpacs,twocolumn]{revtex4}
\usepackage{epsfig}
\usepackage{graphicx,color,dcolumn}
\usepackage{amsmath,amssymb}
\begin{document}
\author{Hui Wang, Zhaozhao Yan, Jialun Ping\footnote{Corresponding author:
J. Ping, jlping@njnu.edu.cn}} \affiliation{Department of Physics,
Nanjing Normal University, Nanjing 210023, P. R. China}

\title{Radially Excited States of $\eta_c$}

\begin{abstract}
In the framework of chiral quark model, the mass spectrum of
$\eta_c(ns)~(n=1,\cdots,6)$ is studied with Gaussian expansion
method. With the wave functions obtained in the study of mass
spectrum, the open flavor two-body strong decay widths are
calculated by using $^3P_0$ model. The results show that the
masses of $\eta_c(1S)$ and $\eta_c(2S)$ are consistent with the
experimental data. The explanation of X(3940) as $\eta_c(3S)$ is
disfavored for X(3940) is a narrow state, $\Gamma=37^{+26}_{-15} \pm 8 $ 
MeV, while the open flavor two-body
strong decay width of $\eta_c(3S)$ is about 200 MeV in our
calculation. Although the mass of X(4160) is about 100 MeV less
than that of $\eta_c(4S)$, the assignment of X(4160) as
$\eta_c(4S)$ can not be excluded because the open flavor two-body
strong decay width of $\eta_c(4S)$ is consistent with the
experimental value of X(4160) and the branching ratios of
$\eta_c(4S)$ are compatible with that of X(4160), and the mass of 
$\eta_c(4S)$ can be shifted downwards by taking into account the
coupling effect of the open charm channels. 
There are still no good candidates to $\eta_c(5S)$ and
$\eta_c(6S)$.
\end{abstract}
\maketitle

\section{Introduction} \label{introduction}

In recent years, a lot of charmonium-like states, so called ''$XYZ$''
states~\cite{XYZ}, have been observed by Belle, BarBar,
BESIII and other collaborations. Most of them cannot be accommodated in
the quark models as conventional mesons because of their exotic properties.
To reveal the underlying
properties of these states has stimulated extensive interest in
the research field of hadron physics.

In the compilation of Particle Data Group (PDG)~\cite{CPC38-090001},
34 states were listed under the $c\bar{c}$ section. Ten states were assigned,
$\eta_c(1S)$, $\eta_c(2S)$, $J/\Psi(1S)$, $\Psi(2S)$, $\chi_{c0}(1P)$,
$\chi_{c0}(2P)$, $\chi_{c1}(1P)$, $\chi_{c2}(1P)$, $\chi_{c2}(2P)$ and $h_c(1P)$,
although there are some controversy about the assignment of
$\chi_{c2}(2P)$~\cite{PRD32-189,PRD72-054026,EPJC72-2226,PRD86-091501R,EPJA50-76}.
Experimentally there is no sign of $X(3915)\rightarrow D\bar{D}$, which strongly
contradicts the theoretical expectation of $\chi_{c2}(2P)$, the $D\bar{D}$ decay
channel should dominate.
In addition, the present analyses strongly favor the following assignments:
$\psi(3770)$ as $1^3D_1$, $\psi(4040)$ as $3^3S_1$, $\psi(4160)$ as $2^3D_1$
and $\psi(4415)$ as $4^3S_1$~\cite{JPCS9,1403.1254}.
The quantum numbers of $X(3872)$ are fixed recently,
$I^G(J^{PC})=0^+(1^{++})$, so it is a good candidate of
$\chi_{c1}(2P)$~\cite{PRL110-222001} although there are also some
arguments about this assignment\cite{X3872-1,X3872-2,X3872-3}.
The explanations of X(3940) as $\eta_c(3S)$~\cite{EPJC74-3208},
X(4140) as $\chi_{c0}(3P)$~\cite{EPJA50-76} are also proposed recently.
However, there are about half of the $c\bar{c}$ states which remain unassigned.

To assign the state reported by experimental collaborations to the theoretical one,
First the masses should be in agreement. Second the decay properties of the states
should be comparable. In the present work, the $\eta_c(nS)~(n=1,\cdots,6)$ states are
studied. The first two states of $\eta_c$ are well established. The other states are
needed to be assigned. To validate the assignment of $\eta_c(ns)~(n=3,\cdots,6)$, the
more rigorous way is to calculate the decay width of these states.
In the present work we study the open charm two-body strong decay
widths of all the $\eta_c(nS)~(n=3,\cdots,6)$ mesons
systematically in a constituent quark model. The spectrum of these
$\eta_c(nS)~(n=1,\cdots,6)$ mesons are obtained by using a
high-precision few-body method, Gaussian expansion method (GEM)
~\cite{GEM}, in the framework of chiral quark
model~\cite{JPG31-481}. In GEM all the interactions are treated
equally rather than some interactions: spin-orbit and tensor
terms, are treated perturbatively in other approaches. The decay
amplitudes to all open charm two-body modes that are nominally
accessible are derived with the $^3P_0$ model. In the numerical evaluation
of the transition matrix elements of decay widths, the wavefunctions obtained
in the study of meson spectrum, rather than the simple harmonic oscillator (SHO) ones,
are used. It is expected to validate the assignment of the radially
excited charmonia with spin-parity $J^{PC}=0^{-+}$ and to provide
useful information for experiment to search the still missing
states.

This work is organized as follows: In section \ref{wavefunctions}, the chiral
quark model and wavefunctions of meson are presented; the $^3P_0$ decay model is
briefly reviewed in section \ref{3p0model}; In section \ref{numcal}, the
numerical results of the two-body decays of $\eta_{c}(nS)(n=3,\cdots,6)$ are
obtained and presented with discussions; And the last section is a short
summary.

\section{The chiral quark model and wave functions}\label{wavefunctions}

The chiral quark model, which has given a good description of hadron
spectra~\cite{JPG31-481,PRC72}, is used to obtain the masses and wavefunctions of $\eta_c$.
Hamiltonian of the model for meson is taken from Ref.~\cite{JPG31-481},
\begin{eqnarray}
 H_{q\bar{q}}(\mathbf{r}) & =& m_1+m_2+\frac{p_{r}^2}{2\mu}+V_{12} \\
 V_{12}& =& V^{C}_{12}+V^{OGE}_{12}+V^{\pi}_{12}+V^{K}_{12}+V^{\eta}_{12}+V^{\sigma}_{12}, \nonumber
\end{eqnarray}
where $m_1$ and $m_2$ are the masses of quark and antiquark, $\mathbf{p}_r$
denotes the relative momentum between quark and antiquark, and $V_{12}$
is the interaction between quark and antiquark. In the present version of the
chiral quark model, the screened color confinement potential is used
\begin{equation}
V_{12}^{C}=\lambda_1 \cdot \lambda_2 \left[ -a_c(1-e^{-\mu_c r})+\Delta \right],
\end{equation}
In some sense, the channel coupling effect of $D\bar{D}$ etc. is taken into account
partly according to Ref.~\cite{PRD79-094004}.
The masses and the wavefunctions of mesons can be obtained by solving the Schr\"{o}dinger
equation,
\begin{equation}
H\Psi_{JM_J}=E \Psi_{JM_J}.
\end{equation}
The wavefunction $\Psi_{JM_J}$ can be written as the direct product of orbital, color,
flavor and spin wavefunctions,
\begin{eqnarray}
\Psi_{JM_J} & =& \left[ \psi_{L}(\mathbf{r})\chi_{S} \right]_{JM_J} \phi(q\bar{q})\omega(q\bar{q}), \\
\left[ \psi_{L}(\mathbf{r})\chi_{S}\right]_{JM_J}  & = & \!\!\!\!
\sum_{M_L,M_S} \!\!\!\!
 \langle LM_LSM_S|JM_J\rangle \psi_{LM_L}(\mathbf{r})\chi_{SM_S},  \nonumber
\end{eqnarray}
where $\langle LM_LSM_S|JM_J\rangle$ is the Clebsh-Gordan
coefficient, $\chi_{SM_S}$, $\phi(q\bar{q})$ and
$\omega(q\bar{q})$ are spin, flavor and color wave function of
meson, respectively. The Gaussian basis functions are employed to expand the orbital wavefunction
$\psi_{LM_L}(\mathbf{r})$~\cite{GEM}
\begin{equation}\label{psilml}
 \psi_{LM_L}(\mathbf{r})=\sum_{k=1}^{k_{max}}C_{Lk}\phi^G_{LMk}(\mathbf{r}).
\end{equation}
\begin{equation}\label{philmk}
 \phi^G_{LMk}(\mathbf{r})=N_{Lk}r^L\exp\left(-\nu_{k}r^2\right)\mathbf{Y}_{LM_L}(\hat{\mathbf{r}}).
\end{equation}
The normalization constant $N_{Lk}$ is
\begin{equation}\label{NLk}
N_{Lk}=\left[\frac{2^{L+2}(2\nu_k)^{L+{\frac{3}{2}}}}{\sqrt{\pi}(2L+1)!!}\right]^{\frac{1}{2}}~~~(k=1,\cdots,k_{max}).
\end{equation}
The Gaussian size parameters are in geometric progression.
\begin{equation}
  \nu_k = \frac{1}{r_{k}^2},~~~~ r_k = r_1 a^{k-1} ~~(k=1,\cdots,k_{max}),
\end{equation}
where $r_1=0.001$ fm, $r_{max}=5.000$ fm and $k_{max}=30$ are used to arrive the
convergent results. Substituting
Eqs.{(4-6) into Eq.(3), we obtain a general eigen-equation,
\begin{equation}
\mathbf{H}\mathbf{c}=E\mathbf{N}\mathbf{c}, \label{ge}
\end{equation}
where $\mathbf{H}$ and $\mathbf{N}$ are hamiltonian and overlap matrices, respectively.

\section{Strong decay and Quark-Pair-Creation model}\label{3p0model}

To calculate the open flavor two-body strong decay widths of
hadrons, the quark-pair-creation model, or $^3P_0$ model, is widely
used. In this model, the hadron decay occurs via a quark-antiquark pair
production from the hadronic vacuum, so the quantum numbers of the created quark
pair are of the hadronic vacuum, $J^{PC}=0^{++}$. This model has given a rather
good description of open flavor two-body strong decay width of
hadrons~\cite{3p0-1,3p0-2,3p0-3,3p0-4,3p0-5,3p0-6}, which are allowed by
Okubo-Zweig-Iizuka (OZI) rule. Here the model is used to calculate the open
charm two-body strong decay widths of the radially excited states of
$\eta_{c}(ns),(n=3,\cdots,6)$. The transition operator used in the model~\cite{3p0-1}
is
\begin{eqnarray}
&&T=-3~\gamma\sum_m\langle 1m1-m|00\rangle\int
d\mathbf{p}_3d\mathbf{p}_4\delta^3(\mathbf{p}_3+\mathbf{p}_4)\nonumber\\
&&~~~~\times{\cal{Y}}^m_1(\frac{\mathbf{p}_3-\mathbf{p}_4}{2})
\chi^{34}_{1-m}\phi^{34}_0\omega^{34}_0b^\dagger_3(\mathbf{p}_3)d^\dagger_4(\mathbf{p}_4).
\end{eqnarray}
The created pair is characterized by a color-singlet wave function $\omega^{34}_0$,
a flavor-singlet function $\phi^{34}_{0}$, a spin-triplet function
$\chi^{34}_{1-m}$ and an orbital wave function
${\cal{Y}}^m_l(\mathbf{p})\equiv|p|^lY^m_l(\theta_p,\phi_p)$ which
is the $l$-th solid spherical harmonic polynomial. $\mathbf{p}_3$
and $\mathbf{p}_4$ denote the momenta carried by the quark and
anti-quark created from the vacuum. The strength of the quark pair creation
$\gamma$ from the vacuum is determined from the
measured partial decay widths. In the present calculation,
$\gamma_n$ and $\gamma_s$ are determined by fitting the open flavor
two-body strong decay widths of the four established states
$\psi(4040)$, $\psi(3770)$ , $\psi(4160)$ and $\chi_{c2}(2P)$ and
the decay widths are showed in Table \ref{fitgamma}. Here
$\gamma_n=4.19$ for $u\bar{u},~d\bar{d}$ pairs and
$\gamma_s=\gamma_n/\sqrt{3}$ for $s\bar{s}$ pair other than that
in Ref.~\cite{PRD72-054026,EPJA50-76}.
\begin{ruledtabular}
\begin{table}[htb]
\caption{The open flavor two-body strong decay widths of
states used to determine $\gamma_n$ and $\gamma_s$. The masses of
mesons involved take the experimental values. (unit in MeV)}
\label{fitgamma}
\begin{tabular}{ccc}
~~~~~Meson~~~~           & $\Gamma$ (exp.)~\cite{CPC38-090001} & $\Gamma$ (theo.)\\
\hline
$\psi(4040)$              &80$\pm$10     &92 \\
$\psi(3770)$            &27.2$\pm$1.0  &16\\
$\psi(4160)$            &70$\pm$10     &56\\
$\chi_{c2}(2P)$          &24$\pm$6      &25 \\

\end{tabular}
\end{table}
\end{ruledtabular}

For the process $A\rightarrow B+C$, the $S$-matrix element is defined
as
\begin{eqnarray}
\langle BC|S|A\rangle=I-2\pi i\delta(E_A-E_B-E_C)\langle BC|T|A\rangle,
\end{eqnarray}
where the T-matrix element is
\begin{eqnarray}
\langle
BC|T|A\rangle=\delta^3(\mathbf{P}_A-\mathbf{P}_B-\mathbf{P}_C){\cal{M}}^{M_{J_A}M_{J_B}M_{J_C}},
\end{eqnarray}
$\mathbf{P}_A$, $\mathbf{P}_B$ and $\mathbf{P}_C$ are the momenta of mesons A ,B and C,
respectively, and ${\cal{M}}^{M_{J_A}M_{J_B}M_{J_C}}$ is the helicity amplitude for the
process $A\rightarrow B + C$.

In experiments, the partial wave decay widths are often used, it can be written as
\begin{eqnarray}
\Gamma = \sum_{JL} \Gamma_{JL}, ~~~~\Gamma_{JL}=\pi^2 \frac{{|\textbf{P}|}}{M_A^2}\Big
|\mathcal{M}^{J L}\Big|^2.\label{partialwidth}
\end{eqnarray}
By the Jacob-Wick formula \cite{AP7-404,AP281-774}, the partial
wave amplitude ${\cal{M}}^{JL}$ can be further related to the
helicity amplitude ${\cal{M}}^{M_{J_A}M_{J_B}M_{J_C}}$,
\begin{eqnarray}
&&{\mathcal{M}}^{J L}(A\rightarrow BC) = \frac{\sqrt{2 L+1}}{2 J_A
+1} \!\! \sum_{M_{J_B},M_{J_C}} \langle L 0 J M_{J_A}|J_A M_{J_A}\rangle \nonumber\\
&&\quad\quad\quad\times\langle J_B M_{J_B}
J_C M_{J_C} | J M_{J_A} \rangle \mathcal{M}^{M_{J_A} M_{J_B}M_{J_C}}({\mathbf{P}}),
\end{eqnarray}
where $\mathbf{J}=\mathbf{J}_B+\mathbf{J}_C$, $\mathbf{J}_{A}
=\mathbf{J}_{B}+\mathbf{J}_C+\mathbf{L}$,
$M_{J_A}=M_{J_B}+M_{J_C}$, and
$\mathbf{P}=\mathbf{P}_B=-\mathbf{P}_C)$ is the three momentum of
the daughter mesons $B$ and $C$ in the center-of-mass frame of
meson A.

\section{Numerical calculation}\label{numcal}
The masses of the involved mesons and the corresponding
wavefunctions are obtained by solving the general eigen-equation
Eq.(\ref{ge}). The running strong coupling constants are taken
from Ref.~\cite{EPJA50-76} and the other parameters are taken
from Ref.~\cite{JPG31-481}. The masses of open-charm mesons and
charmonium $\eta_{c}(ns),(n=1,\cdots,6)$ are shown in Table
\ref{Massofmeson}. For charmonium and most open charm mesons,
there is a good agreement between experimental data and theoretical
results. For several open charm mesons, the theoretical masses
deviate from the experimental data a few percents.

\begin{ruledtabular}
\begin{table}[htb]
\caption{The masses of the open-charm mesons and charmonium $\eta_c$ (unit in MeV).}
\label{Massofmeson}
\begin{tabular}{c|c|c|c|c}
~~~~~Meson~~~~           & GEM & Ref.~\cite{JPG31-481}& Ref.~\cite{PRD72-054026} & Expt.~\cite{CPC38-090001}~~ \\
\hline
$D^0,\bar{D}^0$              &1878  &1883 & - & 1864.84$\pm$0.07 \\
$D^+,D^-$                &1878  &1883 & - & 1869.61$\pm$0.10 \\
$D^{*0},\bar{D}^{*0}$          &2005  &2010 & - & 2006.96$\pm$0.10 \\
$D^{*+},D^{*-}$          &2005  &2010 & - & 2010.26$\pm$0.07 \\
$D^{*0}(2S),\bar{D}^{*0}(2S)$  &2697  & -   & - & - \\
$D^{*+}(2S),D^{*-}(2S)$  &2697  & -   & - & - \\
$D_0^{*0},\bar{D}_0^{*0}$&2431  & -   & - &2318$\pm$29 \\
$D_0^{*+},D_0^{*-}$      &2431  & -   & - &2403$\pm$14$\pm$35 \\
$D_1^{0},\bar{D}_1^{0}$  &2450  &2492 & - &2421.4$\pm$0.6 \\
$D_1^{\pm},\bar{D}_1^{\pm}$  &2450  &2492 & - &2423.2$\pm$2.4 \\
$D_1^{0\prime},\bar{D}_1^{0\prime}$&2529  & -   & - & - \\
$D_2^{*0},\bar{D}_2^{*0}$&2500  &2502 & - &2462.6$\pm$0.6 \\
$D_2^{*\pm},\bar{D}_2^{*\pm}$&2500  &2502 & - &2464.3$\pm$1.6 \\
$D_s^+,D_s^-$           &1968  &1981 & - &1968.30$\pm$0.11\\
$D^{*+}_s,D^{*-}_s$     &2104  &2112 & - &2112.1$\pm$0.4\\
$D^{*+}_{s0},D^{*-}_{s0}$&2460 &2469 & - &2317.7$\pm$0.6\\
$D^{*+}_{s1},D^{*-}_{s1}$&2539 &2543 & - &2459.5$\pm$0.6\\
$D^{*+}_{s1},D^{*-}_{s1}$&2565 &2571 & - &2535.10$\pm$0.08\\
$D^{*+}_{s2},D^{*-}_{s2}$&2583 &2585 & - &2571.9$\pm$0.8\\
$J/\psi(1S)$            &3096  &3097 & 3090 &3096.916$\pm$0.011\\
$\eta_c(1S)$            &2979  &2990 & 2982 &2983.6$\pm$0.7\\
$\psi^{\prime}(2S)$     &3684  &3685 & 3672 &3686.109$^{+0.012}_{-0.014}$ \\
$\eta^{\prime}_c(2S)$   &3622  &3627 & 3630 &3639.4$\pm$1.3\\
$\eta_c(3S)$            &4007  & -   & 4043 &-\\
$\eta_c(4S)$            &4276  & -   & 4384 &-\\
$\eta_c(5S)$            &4470  & -   & -    &-\\
$\eta_c(6S)$            &4612  & -   & -    &-\\
\end{tabular}
\end{table}
\end{ruledtabular}

To justify the assignment, the decay width is very important.
The open charm two-body strong decay modes and decay channels of
$\eta_{c}(nS)(n=3,\cdots,6)$ allowed by the phase space and OZI law
are listed in Table \ref{tabdecaywidth}. The open charm two-body
decay widths of $\eta_{c}(nS)(n=3,\cdots,6)$ are calculated and
also shown in the fourth column of Table \ref{tabdecaywidth}. In
calculating the decay widths, the theoretical masses of mesons
involved and the corresponding wave functions obtained in solving
the Schr\"{o}dinger equations are used. By this way, the
calculation of the widths is more self consistent than most of the
previous works, where the SHO wave functions are used. According
to Ref. \cite{EPJA50-76}, the decay width is sensitive to the
masses of mesons, especially around the threshold of the decay.
For comparison, the results of using experimental masses of mesons in the
calculation of the decay widths are also shown in Table
\ref{tabdecaywidth} (the fifth column).

\begin{ruledtabular}
\begin{table*}[htb]
\caption{The open charm two-body strong decay modes and decay
widths of the possible charmonium states of $\eta_{c}(nS)(n=3,
\cdot, 6)$ allowed by the OZI rule and phase space(unit in MeV).
The widths in the 4th column are derived with the theoretical
masses of the initial state and the final ones. The widths in the
5th column are derived with the experimental masses. And the
widths in the 6th column are from Ref.~\cite{PRD72-054026}.}
\label{tabdecaywidth}
\begin{tabular}{c|c|c|c|c|c}
  State             &Decay mode     &Decay channel             &$\Gamma_a$   &$\Gamma_b$ &Ref.~\cite{PRD72-054026}\\
\hline
  $\eta_c(3S)$   &$0^- + 1^-$    &$D\bar{D}^*,D^*\bar{D},D^{*+}D^-,D^+D^{*-}$  &197.76     &176.18    &47\\
                 &$1^- + 1^-$    &$D^*\bar{D}^*,D^{*+}\bar{D}^{*-}$            &-           &-          &33\\
                                            &&Total                            &197.76    &176.18    &80\\
\hline
 $\eta_c(4S)$   &$0^- + 1^-$    &$D\bar{D}^*,D^*\bar{D},D^{*+}D^-,D^+D^{*-}$   &0.97     & 0.32      &6.3\\
                                &&$D_s^+D_s^{*-},D_s^{*+}D_s^-$                &2.29      & 2.12      &2.2\\
                &$1^- + 1^-$    &$D^*\bar{D}^*,D^{*+}\bar{D}^{*-}$             &68.95     & 71.20      &14\\
                                   &&$D_s^{*+}D_s^{*-}$                        &0.37      & 0.76      &2.2\\
                                                  &&Total                      &72.58   & 74.40      &61\\
\hline
  $\eta_c(5S)$  &$0^- + 1^-$    &$D\bar{D}^*,D^*\bar{D},D^{*+}D^-,D^+D^{*-}$   &6.35      & 7.96      & - \\
                               &&$D_s^+D_s^{*-},D_s^{*+}D_s^-$                 &0.52     & 0.56      & - \\
                &$0^- + 0^+$  &$D\bar{D}_0^{*0},\bar{D}D_0^{*0},D^+D_0^{*-},D^-D_0^{*+}$ &92.22  & 104.89 & - \\
                               &&$D_s^+D^{*-}_{s0},D_s^-D^{*+}_{s0}$           &0.27     & 2.91      & - \\
                &$0^- + 2^+$  &$D\bar{D}_2^{*0},\bar{D}D_2^{*0},D^+\bar{D}_2^{*-},D^-\bar{D}_2^{*+}$ &44.19 & 2.50 & - \\
                &$1^- + 1^-$    &$D^*\bar{D}^*,D^{*+}\bar{D}^{*-}$             &8.69      & 9.99      & - \\
                                   &&$D_s^{*+}D_s^{*-}$                        &0.52      & 0.37      & - \\
                &$1^- + 1^+$ &$D^*\bar{D}_1^{0},\bar{D}^*D_1^{0},D^{*+}D_1^{-},D^{*-}D_1^{+}$ &38.50 & 75.77      & - \\
                                   &&Total                                     &191.26     & 204.95   & - \\
\hline
  $\eta_c(6S)$   &$0^- + 1^-$    &$D\bar{D}^*,D^*\bar{D},D^{*+}D^-,D^+D^{*-}$  &11.48      & 12.77      & - \\
                               &&$D_s^+D_s^{*-},D_s^{*+}D_s^-$                 &0.03       & 0.04     & - \\
                               &&$D\bar{D}^*(2S),D^*\bar{D}(2S),D^{*+}D^-(2S),D^+D^{*-}(2S)$ &1.11  & 5.00  & - \\
                 &$0^- + 0^+$  &$D\bar{D}_0^{*0},\bar{D}D_0^{*0},D^+D_0^{*-},D^-D_0^{*+}$ &6.99  & 2.39 & - \\
                               &&$D_s^+D^{*-}_{s0},D_s^-D^{*+}_{s0}$           &0.66    &1.22      & - \\
                 &$0^- + 2^+$  &$D\bar{D}_2^{*0},\bar{D}D_2^{*0},D^+\bar{D}_2^{*-},D^-\bar{D}_2^{*+}$ &19.41 & 29.13 & - \\
                               &&$D^{+}_sD^{*-}_{s2},D^{-}_sD^{*+}_{s2}$       &0.89     & 0.98     & - \\
                 &$1^- + 1^-$   &$D^*\bar{D}^*,D^{*+}\bar{D}^{*-}$             &0.03   & 0.12     & - \\
                                   &&$D_s^{*+}D_s^{*-}$                        &0.34     & 0.31     & - \\
                 &$1^- + 1^+$ &$D^*\bar{D}_1^{0},\bar{D}^*D_1^{0},D^{*+}D_1^{-},D^{*-}D_1^{+}$ &2.33 &3.26 & - \\
                              &&$D^*\bar{D}_1^{0\prime},\bar{D}^*D_1^{0\prime},D^{*+}D_1^{-\prime},D^{*-}D_1^{+\prime}$ &6.84 & 5.56 & - \\
                              &&$D_s^{*+}D^{*-}_{s1},D_s^{*-}D^{*+}_{s1}$       & -  & 0.14 & - \\
                              &&$D_s^{*+}D^{*-\prime}_{s1},D_s^{*-}D^{*+\prime}_{s1}$ & - & 0.96 & - \\
                 &$1^- + 2^+$ &$D^*\bar{D}_2^{*0},\bar{D}^*D_2^{*0},D^{*+}D_2^{*-},D^{*-}D_2^{*+}$ &2.52 & 0.14  & - \\
                                                           &&Total             &52.63    & 62.02    & - \\

\end{tabular}
\end{table*}
\end{ruledtabular}

From Table \ref{Massofmeson} and \ref{tabdecaywidth}, one can see that
the mass and the open charm two-body decay width of $\eta_{c}(3S)$ are
4007 MeV and 198 MeV, respectively. Around this mass and the quantum numbers
of $\eta_{c}(3S)$ constraint, the possible candidate is X(3940).
By coupling to open charm mesons, the mass of $\eta_{c}(3S)$ can be shifted
downwards. However, the decay width is much higher than the experimental
value of $X(3940)$, $\Gamma=37^{+26}_{-15} \pm 8 $ MeV. And if
the theoretical mass of $\eta_{c}(3S)$ and the experimental masses
of the final states are used, the decay width will reduce to 176
MeV which is still higher than that of $X(3940)$. If the mass of
$\eta_{c}(3S)$ is shifted to 3940 MeV by coupling to $D\bar{D}^*$
channels, the decay width will rise to $\sim 270$ MeV.
In Ref. \cite{PRD72-054026}, the mass of $\eta_{c}(3S)$ is $4043$ MeV
which is $40$ MeV higher than our result. In this case the decay channel to
$D^*\bar{D}^*$ opens and it contributes $33$ MeV to the total
decay width. If the mass of $\eta_{c}(3S)$ rises to $4043$ MeV,
the decay width of $\eta_c(3S)$ to $D\bar{D}^*$ will be $\sim90$
MeV and that of $\eta_c(3S)$ to $D^*\bar{D}^*$ will be $\sim150$
MeV, the total width is over 200 MeV. So the possibility of
assigning $X(3940)$ as $\eta_{c}(3S)$ is not favored in the present work.
In Ref. \cite{EPJC74-3208}, the mass of $\eta_{c}(3S)$ was estimated from
the spectrum pattern and SHO wavefunction was used in the evaluating the
transition matrix element, the decay width of $\eta_{c}(3S)$ is around the
experimental value of $X(3940)$ with appropriate SHO parameter $R$.
So more detailed studies are needed to make the assignment of $\eta_{c}(3S)$.

For $\eta_{c}(4S)$, the mass is 4276 MeV, and the decay width is around
73 MeV. Now the decay of $\eta_{c}(4S)$ to $D^{*}\bar{D}^{*}$ is allowed by
the phase space and it is the main open flavor two-body strong
decay channel. In Ref. \cite{PRD72-054026}, the mass of
$\eta_{c}(4S)$ is $4384$ MeV which is  about $110$ MeV
higher than the result of this work, so that there are more decay modes allowed
by phase space. But its total decay width is not far from our result.
The possible candidate of $\eta_{c}(4S)$ is X(4160),
although its mass is about 110 MeV less than the
theoretical mass of $\eta_c(4S)$. Because the coupling effect of
open charm channels is expected to shift the mass of $\eta_c(4S)$
a little lower. The decay width of X(4160) is 139$^{+111}_{61}\pm 21$ MeV
with $D^*\bar{D}^*$ mode seen and $D\bar{D}$, $D\bar{D}^*+c.c.$ mode not seen.
So the open flavor two-body strong decay width of
$\eta_c(4S)$ is in the range of the experimental value of X(4160),
and the branching ratios of $\eta_c(4S)$ are also compatible with that
of X(4160), where the dacay $\eta_c(4S)\to D\bar{D}$ is forbidden
by the angular momentum coupling and the branching ratio
$\frac{\Gamma(\eta_c(4S))\to
D^*\bar{D}+c.c.)}{\Gamma(\eta_c(4S)\to D^*\bar{D}^*)}=0.014$.
While $\frac{\Gamma(X(4160)\to D\bar{D})}{\Gamma(X(4160)\to
D^*\bar{D}^*)}<0.09$ and $\frac{\Gamma(X(4160)\to
D^*\bar{D}+c.c.)}{\Gamma(X(4160)\to D^*\bar{D}^*)}<0.22$ in
experiment. The assignment of X(4160) as $\eta_{c}(4S)$ cannot be excluded.
This statement is different from the results of Ref. \cite{EPJC74-3208},
where the SHO wavefunctions are used to calculate the transition matrix elements.
For the excited state, the SHO approximation is not reasonable one.

For $\eta_{c}(5S)$ and $\eta_{c}(6S)$ states, the decay to $0^- +
0^+$, $0^- + 2^+$ and $1^- + 1^+$ (even $1^- + 2^+$ for
$\eta_{c}(6S)$) are allowed by the phase space. Moreover they are
the main decay modes of  $\eta_{c}(5S)$ and $\eta_{c}(6S)$ states.
Comparing with experimental data, we cannot find any states with
these properties. Further measurements are expected to identify
these two states.

\section{Summary}\label{shortsummary}
In this work, we study the mass spectra of
$\eta_c(ns)~(n=1,\cdots,6)$ with Gaussian expansion method in the
framework of chiral quark model and calculate the open charm
two-body strong decays of $\eta_c(ns)~(n=3,\cdots,6)$ with $^3P_0$
model. The results show that the masses of $\eta_c(1S)$ and
$\eta_c(2S)$ are consistent with the experimental data. The
explanation of X(3940) as $\eta_c(3S)$ is disfavored because
X(3940) is a narrow state while the open flavor two-body
strong decay width of $\eta_c(3S)$ is about 200 MeV in the present
work. Although the mass of X(4160) is about 100 MeV less than that
of $\eta_c(4S)$, the assignment of X(4160) as $\eta_c(4S)$ can not
be excluded because the coupling effect of open charm channels
may shift the mass of $\eta_c(4S)$ lower, and the open flavor
two-body strong decay width of $\eta_c(4S)$ is in the range of the
experimental value of X(4160) and the branching ratios of
$\eta_c(4S)$ are compatible with that of X(4160).

Because of the opening of open charm decay, the spectra of
$\eta_c(ns)~(n=3,\cdots,6)$ are still not clear as the bottomonium.
To describe the excited spectrum of charmonium, the conventional
quark model needs to be extended. To develop the quark model,
the effect of quark-antiquark pair creation should be taken into
account. For open flavor two-body decay model, $^3P_0$ model,
the improvement is also needed. the dependence of strength $\gamma$ on
the momentum of the created quark has been used to improve the
agreement between theoretical results and experimental data~\cite{NPA683}.
Dynamic model for the meson decay is also expected. The study of the properties of
$\eta_c(ns)~(n=1,\cdots,6)$ is helpful for understanding the
possible exotic, ``$XYZ$" states.

\acknowledgments{ The work is supported partly by the National
Natural Science Foundation of China under Grant Nos. 11035006, 11175088 and 11205091.}


\begin{thebibliography}{99}
\bibitem{XYZ} N. Brambilla, S. Eidelman, B. K. Heltsley, {\em et al.}, Eur. J. Phys. C \textbf{71}, 1534 (2011).
\bibitem{CPC38-090001} K. A. Olive \textit{et al}. [Particle Data Group], Chin. Phys. C \textbf{38}, 090001(2014)
\bibitem{PRD32-189} S. Godfrey and N. Isgur, Phys. Rev. D\textbf{32},189(1985)
\bibitem{PRD72-054026} T. Barnes, S. Godfrey, and E. S. Swanson, Phys. Rev. D \textbf{72},054026(2005).
\bibitem{EPJC72-2226}D. Y. Chen, J. He, X. Liu, \textit{et al}. Eur. Phys. J. C 72:2226-2230(2012) [(arXiv:1207.3561v2 [hep-ph]]
\bibitem{PRD86-091501R} F. K. Guo and ULF-G. Meissner, Phys. Rev. D \textbf{86}, 091501(R) (2012).
\bibitem{EPJA50-76} H. Wang, Y. C. Yang and J. L. Ping, Eur. Phys. J. A \textbf{50}, 76(2014).
\bibitem{1403.1254} S. L. Olsen, arXiv:1403.1254.
\bibitem{JPCS9} T. Barnes, J. Phys. Conf. Ser. \textbf{9}, 127(2005).
\bibitem{PRL110-222001} R. Aaij \textit{et al}. [LHCb Collaboration], Phys. Rev. Lett. \textbf{110}, 222001 (2013).
\bibitem{X3872-1} Y. S. Kalashnikova, Phys. Rev.D {\bf 72}, 034010 (2005). hep-ph/0506270
\bibitem{X3872-2} I. V. Danilkin and Y. A. Simonov, Phys. Rev. Lett. {\bf 105}, 102002 (2010). arXiv:1006.0211[hep-ph]
\bibitem{X3872-3} C. Meng, Y. J. Gao and K. -T. Chao, hep-ph/0506222.
\bibitem{EPJC74-3208} L. P. He, D. Y. Chen, X. Liu and T. Matsuki, Eur. Phys. J. C \textbf{74}, 3208(2014).
\bibitem{GEM} E. Hiyama, Y. Kino, and M. Kamimura, Prog. Part. Nucl. Phys. \textbf{51} 223 (2003).
\bibitem{JPG31-481}J. Vijande , F. Fern¨¡ndez and A. Valcarce, J. Phys. G \textbf{31}, 481(2005).
\bibitem{PRC72} A. Valcarce, H. Garcilazo and J. Vijande, Phys. Rev. C \textbf{72}, 025206 (2005).
\bibitem{PRD79-094004}B. Q. Li and K. T. Chao, Phys. Rev. D\textbf{79},094004(2009)
\bibitem{3p0-1} L. Micu, Nucl. Phys. B {\bf 10}, 521 (1969).
\bibitem{3p0-2} A. Le Yaouanc, L. Oliver, O. Pene, J-C. Raynal, Phys. Rev. D {\bf 8}, 2223 (1973).
\bibitem{3p0-3} A. Le Yaouanc, L. Oliver, O. Pene, J-C. Raynal, Phys. Rev. D {\bf 9}, 1415 (1974).
\bibitem{3p0-4} A. Le Yaouanc, L. Oliver, O. Pene, J-C. Raynal, Phys. Rev. D {\bf 11}, 1272 (1975).
\bibitem{3p0-5} W. Roberts and B. Silvertr-Brac, Few-Body Syst. \textbf{11}, 171 (1992).
\bibitem{3p0-6} E. S. Ackleh, T. Barnes and E. S. Swanson, Phys. Rev. D \textbf{54}, 6811 (1996).
\bibitem{AP7-404} M. Jacob and G. C. Wick, Ann. Phys. (N.Y.) \textbf{7}, 404 (1959).
\bibitem{AP281-774} M. Jacob and G. C. Wick, Ann. Phys. (N.Y.) \textbf{281}, 774 (2000).
\bibitem{NPA683} R. Bonnaz, L. A. Blanco, B. Silvestra, F. Fernandez and A. Valcarce, Nucl. Phys. A \textbf{683}, 425 (2001).
\end{thebibliography}
\end{document}